\documentclass[aps,showpacs,amsmath,amssymb]{revtex4}
\usepackage{graphicx}% Include figure files
\usepackage{dcolumn}% Align table columns on decimal point
\usepackage{bm}% bold math
\usepackage{hyperref}

\newcommand{\beq}{\begin{equation}}
\newcommand{\eeq}{\end{equation}}
\newcommand{\beqa}{\begin{eqnarray}}
\newcommand{\eeqa}{\end{eqnarray}}
\newcommand{\beqar}{\begin{eqnarray*}}
\newcommand{\eeqar}{\end{eqnarray*}}
\newcommand{\bra}[1]{\mbox{$\left\langle{#1}\right|$}}
\newcommand{\ket}[1]{\mbox{$\left|{#1}\right\rangle$}}

\def\I{{\rm i}}

\def\e{{\rm e}}

\newcounter{saveeqn}

\begin{document}

\title{Remote implementations \\
of partially unknown quantum operations of multiqubits}
\author{An Min Wang\footnote{Email Address: anmwang@ustc.edu.cn}}
\affiliation{Quantum Theory Group, Department of Modern Physics\\
University of Science and Technology of China, Hefei 230026, People
Republic of China}

\begin{abstract}

We propose and prove the protocol of remote implementations of
partially unknown quantum operations of multiqubits belonging to the
restricted sets. Moreover, we obtain the general and explicit forms
of restricted sets and present evidence of their uniqueness and
optimization. In addition, our protocol has universal recovery
operations that can enhance the power of remote implementations of
quantum operations.

\end{abstract}

%\keywords{remote implementation of quantum operations, teleportation, quantum communications}

\maketitle

\section{Introduction}

Teleportation of a quantum state \cite{Bennett} means this unknown
state is being transferred from a local system to a remote system
without physically sending the particle. Thus, teleportation of a
quantum operation may be understood as this unknown quantum
operation being transferred from a local system to a remote system
without physically sending the device. However, in the historical
literature, it is more interesting that an unknown quantum operation
acting on the local system (the sender's) is teleported and acts on
an unknown state belonging to the remote system (the receiver's)
\cite{Huelga1}. Taking both teleportation and the action of a
quantum operation into account, one can denote it as ``remote
implementation of operation" (RIO).

If not only a receiver's quantum state (belonging to the remote
system) but also a sender's quantum operation (performing on the
local system) are completely unknown (arbitrary) at the beginning,
the required resource of RIO will be maximum \cite{Huelga1}.
Moreover, if there is a protocol of RIO, then it will be of
significance only when the resource cost of RIO is less than twice
the required resource of teleportation, because that can always be
completed via so-called bidirectional quantum state teleportation
(BQST). Here, BQST contains three steps, that is, the receiver first
teleports an unknown target state to the sender, then the sender
performs an unknown operation (to be remotely implemented) on the
received state to obtain an acted state, and finally the sender
teleports this acted state back to the receiver.

Usually, when a teleported state is partially unknown or partially
known (even completely known), this state transmission process from
a local system to a remote system is called ``remote state
preparation" \cite{Lo, Pati}, while when a teleported operation is
partially unknown or partially known, this operation transmission
process from a local system to a remote system is called ``remote
control of states" \cite{Huelga2}. So-called ``partially unknown" or
``partially known" quantum operations refer to those belonging to
some restricted sets that satisfy some given restricted conditions.
In Ref. \cite{Huelga2}, the authors presented two kinds of
restricted sets of quantum operations in the case of one qubit, that
is, one set consists of diagonal operations and the other set
consists of antidiagonal operations. It is clear that the restricted
sets of quantum operations still include a very large amount of
unitary transformations \cite{Huelga2}. Actually, the remote
implementations of quantum operations belonging to the restricted
sets will consume fewer overall resources than one of completely
unknown quantum operations, and they can satisfy the requirements of
some practical applications. Moreover, the remote implementations of
quantum operations are closely related with nonlocal quantum
operations via local implementations. They both play the important
roles in distributed quantum computation \cite{Cirac,Eisert},
quantum programs \cite{Nielsen,Sorensen} and other tasks of remote
quantum information processing and communication. Recently, a series
of works on the remote implementations of quantum operations
appeared and made some interesting progress both in theory
\cite{Collins,Huelga1,Huelga2} and in experiment
\cite{Huang,Xiang,Huelga3}. Therefore, from our point of view, it is
very important and useful to investigate the extension of remote
implementations of quantum operations to the cases of multiqubits.

To this end, we have to solve some key problems in the cases of
multiqubits, such as how to determine and classify the restricted
sets of quantum operations, how to obtain and express the explicit
form of restricted sets, and finally to present the protocol of
remote implementation of partially unknown quantum operations
belonging to the restricted sets. This paper will focus on these
problems. It must be emphasized that for the cases of $N$ qubits,
the protocol proposed by us only uses $N$ Bell pairs that is half of
the overall quantum resources of the BQST scheme. In addition, there
are universal recovery operations performed by the receiver in this
protocol. This implies that the quantum operations that can be
remotely implemented are extended from within a given restricted set
to all of the restricted sets. One of its advantages is to enhance
the power of remote implementations of quantum operations. This is
useful because one can design the universal recovery quantum
circuits that can be used to the remote implementations of quantum
operations belonging to our restricted sets in the near future.
Because the explicit forms of our restricted sets of multiqubit
quantum operations are not reducible to the direct products of two
restricted sets of one qubit quantum operations, our protocol can be
thought of a development of the scheme of Huelga, Plenio and
Vaccaro's (HPV) \cite{Huelga2}.

This paper is organized as follows. In Sec. \ref{sec2} two, we first
recall HPV protocol and point out its simplification; in Sec.
\ref{sec3}, we obtain the general and explicit form of restricted
sets of $N$ qubit operations, and present evidence of their
uniqueness and optimization in our protocol; in Sec. \ref{sec4}, we
propose the protocol of remote implementations of two-qubit
operations belonging to our restricted sets; in Sec. \ref{sec5}, we
extend our protocol to the cases of $N$ qubits; in Sec. \ref{sec6},
we summarize our conclusions and discuss some problems; in the
appendixes, we explain some notation in this paper, introduce
general swapping transformations, and prove our protocol of remote
implementations of $N$-qubit operations belonging to our restricted
sets.

\section{Simplified HPV protocol}\label{sec2}

The remote implementation of a quantum operation within some given
restricted set was proposed by Huelga, Plenio, and Vaccaro (HPV)
\cite{Huelga2}. In HPV protocol, Alice is set as a sender and Bob is
set as a receiver. Thus, the initial state in the joint system of
Alice and Bob reads \beq \label{inis1}\ket{\Psi_{ABY}^{\rm
ini}}=\ket{\Phi^{+}}_{AB}\otimes\ket{\xi}_Y, \eeq where \beq
\ket{\Phi^{+}}_{AB}=\frac{1}{\sqrt{2}}\left(\ket{00}_{AB}+\ket{11}_{AB}\right)
\eeq is one of four Bell states that are shared by Alice (the first
qubit) and Bob (the second qubit), and the unknown state (the third
qubit) \beq \ket{\xi}_Y=y_0\ket{0}_Y+y_1\ket{1}_Y \eeq belongs to
Bob. Note that Dirac's vectors with the subscripts $A,B,Y$ indicate
their bases, respectively, belonging to the qubits $A,B,Y$.

The quantum operation to be remotely implemented belongs to one of
two restricted sets defined by \beq \label{rs1q}
U(0)=\left(\begin{array}{cc}u_{00}&
0 \\ 0& u_{11}\end{array}\right) \quad U(1)=\left(\begin{array}{cc} 0 & u_{01} \\
u_{10} & 0\end{array} \right).\eeq We can say that they are
partially unknown in the sense that the values of their matrix
elements are unknown, but their structures, that is, the positions
of their nonzero matrix elements, are known. Thus, HPV's protocol
and its simplification can be expressed as the following steps.

{\em Step one}: {\em Bob's preparation}. In the original HPV
protocol, in order to receive the remote control, Bob first performs
a controlled-{\sc not} using his shared part of the $e$-bit as a
control, and then measures his second qubit (the third qubit in the
joint system of Alice and Bob) in the computational bases
$\ket{b}_Y\bra{b}$ $(b=0,1)$. So, Bob's preparation can be written
as \beq {\mathcal{P}}_B^{\rm
original}(b)=\left(\sigma_b^B\otimes\sigma_0^Y\right)
\left(\sigma_0^B\otimes\ket{b}_Y\bra{b}\right)
\left(\ket{0}_B\bra{0}\otimes\sigma_0^Y+
\ket{1}_B\bra{1}\otimes\sigma_1^Y\right),\eeq where $\sigma_0$ is a
$2\times 2$ identity matrix and $\sigma_i$ $(i=1,2,3)$ are the Pauli
matrices. Note that the matrices with the superscripts $A,B,Y$
denote their Hilbert spaces belonging, respectively, to the spaces
of qubits $A,B,Y$. Obviously, the reduced space of Alice or Bob is
easy to obtain by partial tracing.

In fact, the first step in the original HPV protocol can be
simplified by changing Bob's preparation as \cite{Huelga3} \beq
{\mathcal{P}}_B(b)=\left(\ket{b}_B\bra{b}\otimes\sigma_0^Y\right)
\left(\sigma_0^B\otimes\ket{0}_Y\bra{0}
+\sigma_1^B\otimes\ket{1}_Y\bra{1}\right),\eeq that is, Bob first
performs a controlled-{\sc not} using his second qubit (the third
qubit in the joint system of Alice and Bob) as a control, and then
measures his first qubit in the computational bases
$\ket{b}_B\bra{b}$ $(b=0,1)$. This change is very simple but it is
nontrivial because it saves a {\sc not} gate performed by Bob;
moreover, an additional swapping gate at the end of the original HPV
protocol becomes redundant.

{\em Step two}: {\em Classical communication from Bob to Alice}.
After finishing his measurement on the computational basis
$\ket{b}\bra{b}$ $(b=0,1)$, Bob transfers a classical bit $b$ to
Alice. This step is necessary so that Alice can determine her
operation.

It must be emphasized that Bob's preparation can be done in two
equivalent ways with respect to $b=0$ and $1$, respectively. Bob can
fix his measurement as $\ket{0}\bra{0}$ and tells Alice before the
beginning of the protocol, this communication step can be saved, and
then the next Alice's sending step will not need a first $\sigma_b
(=\sigma_0)$ transformation. Similarly, if Bob takes $b=1$ and tells
Alice before the beginning of the protocol, this step can be saved
also, but Alice's next sending step still needs a prior
transformation $\sigma_1$. In the above sense, the protocol may be
able to save a classical bit, even a {\sc not} gate.

{\em Step three}: {\em Alice's sending}. After receiving Bob's
classical bit $b$, Alice first performs a prior transformation
$\sigma_b$ dependent on $b$, and then carries out the quantum
operation $U(d)$ to be remotely implemented on her qubit (the first
qubit). Finally, Alice executes a Hadamard transformation and
measures her qubit in the computational basis $\ket{a}_A\bra{a}$
$(a=0,1)$. All of Alice's local operations and measurement are just
\beq {\mathcal{S}}_A(a,b;d)=\left(\ket{a}_A\bra{a}\right)\left[H^A
U(d)\sigma_b^A\right],\eeq where the Hadamard transformation $H$ is
defined by \beq H=\frac{1}{\sqrt{2}}\left(\begin{array}{cc}1 &1 \\ 1
& -1\end{array}\right). \eeq $U(d)$, defined by Eq. (\ref{rs1q}),
belongs to diagonal or antidiagonal restricted sets, respectively,
when $d=0$ or $1$, and $\sigma_b$ is taken as $\sigma_0$- or
$\sigma_1$-dependent on the received classical information $b=0$ or
$1$.

{\em Step four}: {\em Classical communication from Alice to Bob}.
After finishing her measurement on the computational basis
$\ket{a}_A\bra{a}$ $(a=0,1)$, Alice transfers a classical bit $a$ to
Bob. Moreover, Alice also needs to transfer an additional classical
information $d=0$ or $1$ in order to tell Bob whether the
transferred operation is diagonal or antidiagonal, unless they have
prescribed the transferred operation belonging to a given restricted
set before the beginning of the protocol.

{\em Step five}: {\em Bob's recovery}. In order to obtain the remote
implementation of this quantum operation in a faithful and
determined way, Bob has to preform his recovery operation in
general. In the original HPV protocol, this operation is \beq
{\mathcal{R}}_B^{\rm original}(a;d)=\left\{\left[(1-a)
\sigma_0^B+a\sigma_3^B\right]\sigma_d^B\right\}\otimes\sigma_0^Y.\eeq
In the simplified HPV protocol, Bob's recovery operation becomes
\beq {\mathcal{R}}_B(a;d)=\sigma_0^B\otimes\left\{\left[(1-a)
\sigma_0^Y+a\sigma_3^Y\right]\sigma_d^Y\right\}.\eeq

It is clear that the original HPV protocol will result in
$U(d)\left(y_0\ket{0}_B+y_1\ket{1}_B\right)$ in the second qubit of
the joint system. One cannot help to perform an additional swapping
operation between the second qubit and the third qubit defined by
\beq {\mathcal{B}}_{\rm swap}^{\rm original}=
\left(\begin{array}{cccc}1&0&0&0\\0&0&1&0\\0&1&0&0\\0&0&0&1\end{array}\right).
\eeq

However, in terms of the simplified HPV protocol, after carrying out
the above steps from one to five, we can directly obtain
$U(d)\left(y_0\ket{0}_Y+y_1\ket{1}_Y\right)$ in the third qubit of
the joint system. This means that an additional swapping step has
been saved.

All of the operations including measurements in the simplified HPV
protocol can be jointly written as \beq {\mathcal{I}}_R(a,b;d)=
\left[\sigma_0^A\otimes{\mathcal{R}}_B(a;d)\right]
\left[{\mathcal{S}}_A(a,b;d)\otimes\sigma_0^B\otimes\sigma_0^Y\right]
\left[\sigma_0^A\otimes{\mathcal{P}}_B(b)\right]. \eeq Its action on
the initial state (\ref{inis1}) gives \beq \ket{\Psi_{ABY}^{\rm
final}(a,b;d)}={\mathcal{I}}_R(a,b;d)\ket{\Psi_{ABY}^{\rm ini}}=
\frac{1}{2}\ket{a b}_{AB}\otimes U(d)\ket{\xi}_Y. \eeq where $a,b=0$
or $1$ denotes the spin up or spin down, and $d=0$ or $1$ indicates
the diagonal operation or antidiagonal operation, respectively.
Therefore, the remote implementations of one-qubit quantum
operations belonging to two restricted sets are faithfully and
determinedly completed.

It is easy to plot the quantum circuit of the simplified HPV
protocol; see Fig. 1.

\begin{figure}[ht]%[tbp]
\begin{center}
\includegraphics[scale=0.60]{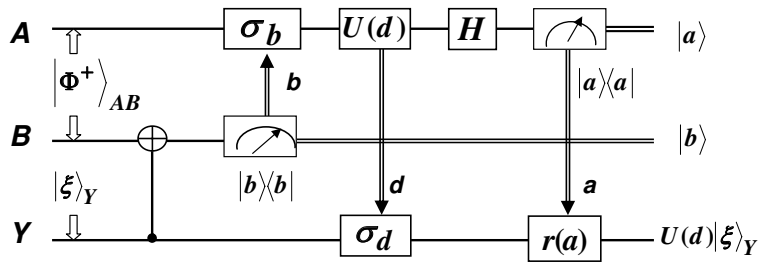}
\end{center}
\vskip -0.1in \caption{Quantum circuit of the simplified HPV
protocol, where $U(d)$ is a quantum operation to be remotely
implemented and it is diagonal or antidiagonal, $H$ is a Hadamard
gate, $\sigma_b,\sigma_d$ are identity matrices or {\sc not} gates
($\sigma_1$) with respect to $b,d=0$ or $b,d=1$, respectively, and
$r(a)=(1-a)\sigma_0+a\sigma_3$ is an identity matrix when $a=0$ or a
phase gate ($\sigma_3$) when $a=1$. The measurements
$\ket{a}\bra{a}$ and $\ket{b}\bra{b}$ are carried out in the
computational basis ($a,b=0,1$). ``$\Rightarrow$" (crewel with an
arrow) indicates the transmission of classical communication to the
location of the arrow direction.} \label{mypic1}
\end{figure}

\section{Restricted sets of quantum operations}\label{sec3}

We have described the simplified HPV protocol of remote
implementations of one qubit quantum operations in detail. For our
purpose to extend it to the cases of multiqubits, we first seek for
the restricted sets of multiqubit quantum operations that can be
remotely implemented in a faithful and determined way. Here, through
analyzing and discussing the cases of one- and two-qubit operations,
we can exhibit our method to obtain the general and explicit forms
of restricted sets of multiqubit quantum operations.

Let us start with the analysis of HPV protocol for one qubit. From
our point of view, the purpose of Bob's preparation is to lead to
the first qubit (locally acted qubit in Alice's subsystem) being
correlated with the third qubit (remotely operated or controlled
qubit in Bob's subsystem) in such a superposition that for its every
orthogonal component state, the first qubit and the third qubit are
always located at the same computational bases. Bob arrives at this
aim with two possible ways via quantum entanglement resource between
the first qubit and the second qubit (in Bob's subsystem). When Bob
uses $b=0$, then this aim has been achieved at, but if Bob takes
$b=1$, Alice has to supplement a $\sigma_1$ transformation for this
aim. It is clear that such a superposition state has at most two
orthogonal component states that is equal to the dimension of
Hilbert's space of an unknown state. This implies that we can, at
most, transfer two unknown complex numbers from the first qubit to
the third qubit. We think that this is a really physical reason why
we can only remotely implement a quantum operation belongings to the
restricted sets. Without using additional correlation
(entanglement), we cannot change this physical fact. However, using
additional entanglement will destroy our attempt to save quantum
resources.

In the second step, the communication from Bob to Alice is to tell
Alice which preparing way Bob has used. In order to include all
contributions of operation on the first qubit and transfer them to
the third qubit, we need a Hadamard gate acting on the transformed
qubit so that Alice's project measurement on a given computational
basis does not lead to losing the actions on the other computational
bases, because the first qubit and the third qubit are correlated in
the above way. However, the action of the Hadamard gate will result
in an algebraic addition of all of matrix elements in some row or
column of this operation arising in front of some computation bases.
Its advantage is that we are able to transfer the whole effect of
operation to the third qubit, but its disadvantage is that we are
not able to redivide the algebraic addition of matrix elements in
some row or column of this operation because these elements are
unknown. A uniquely choice way is to set only one nonzero element in
every row or every column of this operation. In fact, this choice is
also optimal since it allows the maximal numbers that can be
transferred and also includes the unitary operations with such
forms. This requirement yields the limitations to the structures of
operations that can be remotely implemented, that is, so-called
restricted sets of quantum operations. In the case of one qubit, it
is easy to see that two restricted sets of quantum operations are
made from a kind of diagonal operation and a kind of antidiagonal
operation.

For the cases of two qubits, the above analyses are still feasible
and valid. Because the unique nonzero element in the first row has
four possible positions, the unique nonzero element in the second
row has three possible positions, the unique nonzero element in the
third row has two possible positions, and the unique nonzero element
in the fourth row has one possible position, the restricted sets of
operations are made of $4!=24$ kinds of operations.

It is easy to write the set of all of permutations for the list
$\{1,2,3,4\}$, \beqa \mathbb{P}_4&=& \{(1, 2, 3, 4), (1, 2, 4, 3),
(1, 3, 2, 4), (1, 3, 4, 2), (1, 4, 2, 3), (1, 4,3, 2),\nonumber\\ &
&(2, 1, 3, 4), (2, 1, 4, 3), (2, 3, 1, 4), (2, 3, 4, 1), (2, 4, 1,
3), (2, 4, 3, 1),\nonumber\\ & &(3, 1, 2, 4), (3, 1, 4, 2), (3, 2,
1, 4), (3, 2, 4, 1), (3, 4, 1, 2), (3, 4, 2, 1),\nonumber\\ & &(4,
1, 2, 3), (4, 1, 3, 2), (4, 2, 1, 3), (4, 2, 3, 1), (4, 3, 1, 2),
(4, 3, 2, 1)\}. \eeqa Denoting the $x$th element in this set by \beq
p(x)=(p_1(x),p_2(x),p_3(x),p_4(x)),\eeq for example $p(1)=(1,2,3,4),
p(2)=(1,2,4,3)$, and so on, we can obtain 24 restricted sets of
two-qubit operations as follows: \beq
T_2^r(x,t)=\sum_{m=1}^4t_m\ket{m,D}\bra{p_m(x),D}, \eeq where we
have defined $\ket{1,D}=\ket{00},
\ket{2,D}=\ket{01},\ket{3,D}=\ket{10},\ket{4,D}=\ket{11}$. Here, the
label $D$ indicates the decimal system.

It is easy to verify that \beqa
T_2^r(x,t)[T_2^r(x,t)]^\dagger &=&\sum_{m=1}^4t_mt_m^*\ket{m,D}\bra{m,D} \\
\left[T_2^r(x)\right]^\dagger T_2^r(x)&=&\sum_{m=1}^4t_it_m^*
\ket{p_m(x),D}\bra{p_m(x),D}.\eeqa Therefore, in terms of the
requirement of the unitary condition for quantum operations, the
only nonzero element $t_m$ in the $m$th row of quantum operations
belonging to the restricted sets should be taken as $\e^{\I\phi_m}$,
and $\phi_m$ is real.

The above analyses and discussions have provided evidence of unique
forms of restricted sets of two-qubit operations in a kind of
protocol of RIO such as ours. In fact, this kind of protocol uses
the Hadamard gates to transfer the whole effect of operation to the
different qubits, but does not use the extra correlation doing it.
Therefore, the forms of restricted sets are uniquely determined.
Otherwise, the operation cannot be remotely implemented by using
such a kind of protocol.

To remotely implement quantum operations belonging to the above
restricted sets, Bob needs a mapping table that provides one-to-one
mapping from a classical information $x$ $(x=1,2\cdots,24)$ to a
part of his recovery operation $R_2(x)$ defined by \beq
\label{fixedu}{R}_2(x)=
T_2^r(x,0)=\sum_{m=1}^4\ket{m,D}\bra{p_m(z),D}.\eeq Obviously, it
has the same structure as $T_2^r(x,t)$ to be remotely implemented.

It is easy to see that the controlled kinds of operations
\begin{gather}\label{co1}U_C(1)=\left.T_2^r(2,t)\right|_{t_1=t_2=1}=\begin{pmatrix}1& 0 \\ 0&0\end{pmatrix}
\otimes\begin{pmatrix}1& 0 \\ 0&1\end{pmatrix} + \begin{pmatrix}0& 0
\\ 0&1\end{pmatrix} \otimes\begin{pmatrix}0& t_3 \\ t_4&0\end{pmatrix}
\end{gather}
\begin{gather}\label{co2}
U_C(2)=\left.T_2^r(6,t)\right|_{t_1=t_3=1}=\begin{pmatrix}1& 0 \\
0&1\end{pmatrix} \otimes\begin{pmatrix}1& 0 \\ 0&0\end{pmatrix}
+\begin{pmatrix}0 &
t_2 \\ t_4&0 \end{pmatrix} \otimes\begin{pmatrix}0 & 0 \\
0&1\end{pmatrix}
\end{gather}
\begin{gather}\label{co3}U_C(3)=\left.T_2^r(7,t)\right|_{t_1=t_2=1}= \begin{pmatrix}1& 0
\\ 0&0\end{pmatrix} \otimes\begin{pmatrix}0& t_3 \\
t_4&0\end{pmatrix}+\begin{pmatrix}0& 0 \\ 0&1\end{pmatrix}
\otimes\begin{pmatrix}1& 0 \\ 0&1\end{pmatrix}
\end{gather}
\begin{gather}\label{co4}
U_C(4)=\left.T_2^r(15,t)\right|_{t_2=t_4=1}=\begin{pmatrix}0 &
t_1 \\ t_3&0 \end{pmatrix} \otimes\begin{pmatrix}1 & 0 \\
0&0\end{pmatrix}+\begin{pmatrix}1& 0 \\ 0&1\end{pmatrix}
\otimes\begin{pmatrix}0& 0 \\ 0&1\end{pmatrix}
\end{gather}
belong to the restricted set. They are important operations in
quantum information processing.

Based on the same reasons stated above, any restricted set of
$N$-qubit operations has such a structure that every row and every
column of its operations only has one nonzero element, and we denote
this nonzero element in the $m$th row by $t_m$, that is, the members
of $2^N!$ restricted sets of $N$-qubit operations have the forms
\beq T^r_N(x,t)=\sum_{m=1}^{2^N}t_m\ket{m,D}\bra{p_m(x),D},\eeq
where $x=1,2,\cdots,2^N!$ and \beq
p(x)=(p_1(x),p_2(x),\cdots,p_{2^N}(x))\eeq is an element belonging
to the set of all permutations for the list $\{1,2,\cdots,2^N\}$.
All of the restricted sets of $N$-qubit operations are denoted by
$\mathbb{T}^r_N$.

For the cases of $N$-qubit operations, we can take all nonzero
elements of $T^r_N(x,t)$ as $1$ and obtain its fixed form $R_N(x)$,
that is \beq
{R}_N(x)=T^r_N(x,0)=\sum_{m=1}^{2^N}\ket{m,D}\bra{p_m(x),D}.\eeq It
will be used in Bob's recovery operation of our protocol.

It must be emphasized that we usually study the cases in which
$T^r_N(x,t)$ is unitary, although it does not affect our protocol.
Before the beginning of the protocol, we need to build two mapping
tables: one provides one-to-one mapping from $T^r_N(x,t)\in
\mathbb{T}^r_N$ to the classical information $x$ which is known by
Alice and another provides one-to-one mapping from a classical
information $x$ to $R_N(x)$ which is known by Bob.

It is clear that our explicit restricted sets of multiqubit
operations are not reducible to the simple direct product of two
restricted sets of one qubit operations. Thus, in this sense, our
protocol can be thought of as a development of HPV protocol to the
cases of multiqubits.

\section{Protocol in the case of two qubits}\label{sec4}

Now let us propose the protocol of remote implementations of
two-qubit quantum operations belonging to 24 restricted sets in
detail.

Assume the initial state of the joint system to be \beq
\label{inis2}\ket{\Psi_{A_1B_1A_2B_2Y_1Y_2}^{\rm
ini}}=\ket{\Phi^{+}}_{A_1B_1}\otimes\ket{\Phi^{+}}_{A_2B_2}\ket{\xi}_{Y_1Y_2},
\eeq where the unknown state of two qubits is \beq
\ket{\xi}_{Y_1Y_2}=\sum_{j_1,j_2=0}^1
y_{j_1j_2}\ket{j_1j_2}_{Y_1Y_2}, \eeq the qubits $A_1,A_2$ belong to
Alice, the other four qubits $B_1,B_2,Y_1,Y_2$ are owned by Bob. It
is clear that Alice and Bob share initially two Bell states.

Note that the Hilbert space of the joint system is initially taken
as a series of direct products of Hilbert spaces of all qubits
according to the following sequence: \beq H=H_{A_1}\otimes
H_{B_1}\otimes H_{A_2}\otimes H_{B_2}\otimes H_{Y_1}\otimes
H_{Y_2}.\eeq We can simply call this sequence ``space structure" and
denote it by a bit-string; for example, the space structure of the
above Hilbert space is $A_1B_1A_2B_2Y_1Y_2$. Obviously, taking such
a space structure, the subspace belonging to Alice or Bob is
separated. It will lead to inconvenience in the expression of local
operations acting on their full subspaces and in the proof of the
protocol of multiqubits. Therefore, there is a need to change the
space structure. This can be realized by a series of swapping
transformations, which are studied in Appendix A.

In terms of the general swapping transformations defined in Appendix
A, we can change the initial space structure, for example, \beqa
\label{inist1}\ket{a_1b_1a_2b_2y_1y_2}_{A_1B_1A_2B_2Y_1Y_2}
&=&\Upsilon^{-1}(3,2)\left(\ket{a_1b_1y_1}_{A_1B_1Y_1}
\otimes\ket{a_2b_2y_2}_{A_2B_2Y_2}\right)\\
\label{inist2}\ket{a_1b_1a_2b_2y_1y_2}_{A_1B_1A_2B_2Y_1Y_2}
&=&\left[\Lambda^{-1}(2,2)\otimes
I_4\right]\left(\ket{a_1a_2}_{A_1A_2}\otimes
\ket{b_1b_2}_{B_1B_2}\otimes\ket{y_1y_2}_{Y_1Y_2}\right)\\
\label{inist3}\ket{a_1b_1a_2b_2y_1y_2}_{A_1B_1A_2B_2Y_1Y_2}
&=&\Gamma^{-1}(3,2)\left(\ket{a_1a_2}_{A_1A_2}
\otimes\ket{y_1y_2}_{Y_1Y_2}\otimes\ket{b_1b_2}_{B_1B_2}\right).
\eeqa Thus, we can express our formula compactly and clearly in the
whole space, and can finally prove our protocol conveniently and
strictly. Our notations in the whole space will be helpful for in
understanding the problems even if a little complication in
expressions is induced. It will be seen that such notations are more
useful for the extension to the cases of multiqubits. However, it
must be emphasized that these swapping transformations in the
following formula do not really exist in the practical process.

{\em Step One}: {\bf Bob's preparation}. Our protocol begins from
this step. Bob first performs two controlled-{\sc not} using,
respectively, his qubits $Y_1$ and $Y_2$ as two control qubits,
$B_1$ and $B_2$ as two target qubits, and then measures his two
qubits $B_1$ and $B_2$ in the computational basis
$\ket{b_1}_{B_1}\bra{b_1}\otimes\ket{b_2}_{B_2}\bra{b_2}$
$(b_1,b_2=0,1)$. Therefore, Bob's preparation reads \beq
{\mathcal{P}}_B(b_1,b_2)=\Upsilon^{-1}(3,2)\left\{\bigotimes_{m=1}^2\sigma_0^{A_m}
\otimes\left[\left(\ket{b_m}_{B_m}\bra{b_m}\otimes
\sigma_0^{Y_m}\right)C^{\rm
not}(0,1)\right]\right\}\Upsilon(3,2),\eeq where $\Upsilon(3,N)$ is
defined in Appendix A. Note that this expression is written in the
whole joint system so that we can prove our protocol more
conveniently in Appendix B.

If we do not use the swapping transformations, the form of Bob's
preparation becomes \beqa
{\mathcal{P}}_B(b_1,b_2)&=&\left(\sigma_0^{A_1}\otimes\ket{b_1}_{B_1}\bra{b_1}
\otimes\sigma_0^{A_2}\otimes\ket{b_2}_{B_2}\bra{b_2}
\otimes\sigma_0^{Y_1}\otimes\sigma_0^{Y_2}\right)\cdot\left(\sigma_0^{A_1}\otimes
C^{\rm
not}_2(0,1)\otimes\sigma_0^{Y_2}\right)\nonumber\\
& &\times\left(\sigma_0^{A_1}
\otimes\sigma_0^{B_1}\otimes\sigma_0^{A_2}\otimes C_1^{\rm
not}(0,1)\right). \eeqa Here, $C_{M}^{\rm not}$ can be called the
separated controlled-{\sc not} since its control and target are
separated by $M$ qubits, that is, its definition is
\beq\label{separatecnot} C^{\rm
not}_M(0,1)=\sigma_0\otimes\left(\bigotimes_{m=1}^M
\sigma_0\right)\otimes\left(\ket{0}\bra{0}\right)
+\sigma_1\otimes\left(\bigotimes_{m=1}^M\sigma_0\right)
\otimes\left(\ket{1}\bra{1}\right), \eeq while $(0,1)$ indicates
that the last qubit is a control and the first qubit is a target and
is flipped when the control qubit is $\ket{1}$. If $M=0$, it comes
back to the usual controlled-{\sc not}.  It is clear that using the
general swapping transformations can simplify the expressions of
formula in form.

{\em Step Two}: {\bf Classical communication from Bob to Alice}.
After finishing his measurement on the computational basis, Bob
transfers two classical bits $b_1,b_2$ to Alice. This step is
necessary so that Alice can determine her sending operations.

It must be emphasized that Bob's preparation step has four
equivalent ways corresponding to, respectively, $b_1b_2$ taking
$00,01,10,11$ in order to carry out the protocol. If Bob first fixes
the value of $b_1b_2$ and tells Alice before the beginning of the
protocol, this step can be saved. In particular, when $b_1b_2$ is
just taken as $00$, Alice also does not need the transformation
$\sigma_{b_1}\otimes\sigma_{b_2}$ in the next step, since
$\sigma_{0}\otimes\sigma_{0}$ is trivial.

{\em Step Three}: {\bf Alice's sending}. After receiving Bob's
classical bits $b_1b_2$, Alice, on her two qubits (the qubits
$A_1A_2$), first performs $\sigma_{b_1}^{A_1}
\otimes\sigma_{b_2}^{A_2}$, secondly acts $T_2^r(x,t)$ to be
remotely implemented, then carries out two Hadamard transformations,
and finally measures her two qubits in the computational basis
$\ket{a_1}_{A_1}\bra{a_1}\otimes \ket{a_2}_{A_2}\bra{a_2}$
$(a_1,a_2=0,1)$.  Since the basis vector of Alice's space has the
structure $\ket{a_1a_2}_{A_1A_2}$, all of Alice's local operations
and measurement are just \beqa
{\mathcal{S}}_A(a_1,b_1,a_2,b_2;x,t)&=&\left[\Lambda^{-1}(2,2)\otimes
I_4\right]\left\{\left[ \left(\ket{a_1a_2}_{A_1A_2}\bra{a_1a_2}
\right)\left(H^{A_1}\otimes
H^{A_2}\right)\right.\right.\nonumber\\
& &\left.\left.\times T_2^r(x,t)\left(\sigma_{b_1}^{A_1}
\otimes\sigma_{b_2}^{A_2}\right)\right]\otimes
I_{16}\right\}\left[\Lambda(2,2)\otimes I_{4}\right], \eeqa where
$\Lambda(2,2)$ is defined in Appendix A and $I_m$ is a $m$
dimensional identity matrix.

{\em Step Four}: {\bf Classical communication from Alice to Bob}.
After finishing her measurement on the computational basis
$\ket{a_1}_{A_1}\bra{a_1}\otimes \ket{a_2}_{A_2}\bra{a_2}$
$(a_1,a_2=0,1)$, Alice transfers two classical bits $a_1,a_2$ to
Bob. Moreover, Alice also needs to transfer $x$ (which can be
encoded by five classical bits) to Bob in order to let him know the
transferred operation $T_2^r(x,t)$ belonging to which restricted
set, unless they prescribed the transferred operation $T_2^r(x,t)$
belonging to a given restricted set before the beginning of the
protocol. All of the classical information is necessary for Bob so
that he can determine his recovery operations.

{\em Step Five}: {\bf Bob's recovery}. In order to obtain the remote
implementations of quantum operations in a faithful and determined
way, Bob performs his recovery operation, \beq
{\mathcal{R}}_B(a_1,a_2;x)=I_{16}\otimes
\left\{\left[\mathfrak{r}^{Y_1}(a_1)\otimes\mathfrak{r}^{Y_2}(a_2)\right]\cdot
{R}_2(x)\right\},\eeq where $\mathfrak{r}(y)$ is defined by \beq
\mathfrak{r}(y)=(1-y) \sigma_0+y\sigma_3,\eeq while $R_2(x)$ is
obtained by the mapping table from the classical information $x$ to
$R_2(x)$. For example, Bob receives $1$ (which can be encoded by
$00000$), thus he knows $R(1)$ is an identity matrix; Bob receives
$2$ (which can be encoded by $00001$), thus he
knows \beq R_2(2)=\left(\begin{array}{cccc}1&0&0& 0\\
0&1& 0 & 0\\0&0&0&1\\0&0&1&0\end{array}\right),\eeq and so on. The
mapping between $x$ and $R_2(x)$ is given in advance before the
beginning of the protocol.

Finally, all of the operations including measurements in the whole
space for the remote implementations of quantum operations of two
qubit can be written jointly as \beq
{\mathcal{I}}_R(a_1,b_1,a_2,b_2;x,t)={\mathcal{R}}_B(a_1,a_2;x)\cdot
{\mathcal{S}}_A(a_1,b_1,a_2,b_2;x,t)\cdot
{\mathcal{P}}_B(b_1,b_2).\eeq Its action on the initial state gives
the remote implementations of two qubit quantum operations belonging
to the above 24 restricted sets, that is, the final state becomes
\beqa \label{rio2q} \ket{\Psi_{A_1B_1A_2B_2Y_1Y_2}^{\rm
final}(a_1,b_1,a_2,b_2;x,t)} &=&
{\mathcal{I}}_R(a_1,b_1,a_2,b_2;x)\ket{\Psi_{A_1B_1A_2B_2Y_1Y_2}^{\rm
ini}}\\&=& \frac{1}{4}\ket{a_1 b_1 a_2 b_2}_{A_1B_1A_2B_2}\otimes
T_2^r(x,t)\ket{\xi}_{Y_1Y_2}, \eeqa where $a_m,b_n=0,1$; $m,n=1,2$.
Therefore, our protocol completes faithfully and determinedly the
remote implementations of quantum operations $T_2^r(x,t)$ belonging
to 24 restricted sets. Its proof is found in Appendix B when $N=2$.

\section{Extension to the cases of $N$ qubits}\label{sec5}

Based on our above protocol of remote implementations of two-qubit
operations belonging to our restricted sets, we can extend it to the
cases of more than two qubits without obvious difficulty. Our
protocol consists of five steps for the remote implementations of
$N$-qubit operations belonging to our restricted sets. Set the
initial state as \beq \ket{\Psi_N^{\rm
ini}}=\left(\bigotimes_{m=1}^N\ket{\Phi^+}_{A_mB_m}\right)
\otimes\ket{\xi}_{Y_1Y_2\cdots Y_N}, \eeq where
$\ket{\xi}_{Y_1Y_2\cdots Y_N}$ is an arbitrary (unknown) pure state
in an $N$-qubit system, that is \beq \ket{\xi}_{Y_1Y_2\cdots
Y_N}=\sum_{k_1,k_2,\cdots k_N=0}^1 y_{k_1k_2\cdots
k_N}\ket{k_1k_2\cdots k_N}.\eeq It is clear that the space structure
is initially \beq \prod_{m=1}^N(A_mB_m)\prod_{n=1}^NY_n.\eeq

Usually, in order to avoid possible errors and provide convenience
in the proof, we need to set the sequential structure of direct
product space of qubits, or a sequence of direct products of
qubit-space basis vectors in the multiqubit systems. For Alice's
space, we set its sequential structure as $A_1A_2\cdots A_N$, in
other words, its basis vector has the form
$\ket{a_1}_{A_1}\ket{a_2}_{A_2}\cdots\ket{a_N}_{A_N}$ (or
$\ket{a_1a_2\cdots a_N}_{A_1A_2\cdots A_N}$). Similarly, we set the
sequential structure of Bob's space as $B_1B_2\cdots B_NY_1Y_2\cdots
Y_N$, in other words, its basis vector has the form
$\ket{b_1}_{B_1}\ket{b_2}_{B_2}\cdots\ket{b_N}_{B_N}
\ket{y_1}_{Y_1}\ket{y_2}_{Y_2}\cdots\ket{y_N}_{Y_N}$. It is clear
that for an $N$-qubit system, its space structure can be represented
by a bit-string with the length of $N$.

Now, let us describe our protocol in a concise way.

{\em Step One}: {\bf Bob's preparation}

\beq {\mathcal{P}}_B(b_1,b_2,\cdots,b_N)=\Upsilon^{-1}(3,N)
\left\{\bigotimes_{m=1}^N\sigma_0^{A_m}\otimes
\left[\left(\ket{b_m}\bra{b_m}\otimes\sigma_0\right) C^{\rm
not}(0,1)\right]\right\}\Upsilon_N(3,N), \eeq where $\Upsilon(3,N)$
is defined in Appendix A. It must be emphasized that $\Upsilon(3,N)$
does not appear in the practical process, it is only required to
express our steps clearly and compactly.

{\em Step Two}: {\bf Classical Communication from Bob to Alice}.
Alice transfers a classical bit-string $b_1b_2\cdots b_N$ to Bob
unless Bob and Alice have an arrangement about Bob's preparing
method (that is $b_1b_2\cdots b_N$ to be determined by Bob and known
by Alice) before the beginning of the protocol.

{\em Step Three}: {\bf Alice's sending}. \beqa
{\mathcal{S}}_A(a_1,b_1,a_2,b_2,\cdots,a_N,b_N;x,t)
&=&\left(\Lambda^{-1}(2,N)\otimes
I_{2^N}\right)\left[\left(\bigotimes_{m=1}^N
\ket{a_m}_{A_m}\bra{a_m}\right)\cdot \left(\bigotimes_{m=1}^N
H^{A_m}\right)\right.\nonumber\\
& &\left. \cdot
T^r_N(x,t)\cdot\left(\bigotimes_{m=1}^N\sigma_{b_m}^{A_m}\right)\otimes
I_{4^N}\right]\left(\Lambda(2,N)\otimes I_{2^N}\right),\eeqa where
$\Lambda_N(2,N)$ is defined in Appendix A.

{\em Step Four}: {\bf Classical Communication from Alice to Bob}.
Alice transfers a classical bit-string $a_1a_2\cdots a_N$ and a
classical information $x$ (which can be encoded by
$\left[\log_2(2^N!)\right]+1$ $c$-bit string, where $[\cdots]$ means
taking the integer part) corresponding to the quantum operation
$T^r_N(x,t)$ to be remotely implemented in her mapping table.

{\em Step Five}: {\bf Bob's recovery}. \beq
{\mathcal{R}}_B(a_1,a_2\cdots
a_N;x)=I_{4^N}\otimes\left\{\!\!\left(\bigotimes_{m=1}^N
\mathfrak{r}(a_m)\right)\cdot {R}_N(x)\right\},\eeq where $R_N(x)$
is determined by Bob's mapping table.

Thus, all of the operations including measurements in the extension
of remote implementations of quantum operations to the case of $N$
qubits can be written as \beqa
{\mathcal{I}}_R(a_1,b_1,a_2,b_2;\cdots,a_N,b_N;x,t)
&=&{\mathcal{R}}_B(a_1,a_2,\cdots a_N;x)\nonumber\\
& &\times {\mathcal{S}}_A(a_1,b_1,a_2,b_2,\cdots, a_N,b_N;x,t)\nonumber\\
& &\times {\mathcal{P}}_B(b_1,b_2,\cdots,b_N).\eeqa

The final state becomes \beqa & &\ket{\Psi_N^{\rm
final}(a_1,b_1,a_2,b_2,\cdots,a_N,b_N;x)}\nonumber\\& &\quad =
{\mathcal{I}}_R(a_1,b_1,a_2,b_2;\cdots,a_N,b_N;x,t)\ket{\Psi_N^{\rm
ini}}
\\ & &\quad = \frac{1}{2^N}\left(\bigotimes_{i=1}^N\ket{a_i b_i}_{A_iB_i}\right)\otimes
T^r_N(x,t)\ket{\xi}_{Y_1Y_2\cdots Y_N}. \eeqa where $a_m,b_n=0,1$;
$m,n=1,2,\cdots, N$.

It is easy to see that our restricted sets of three-qubit operations
include the interesting controlled-controlled-$U(d)$ gate with the
form \beq U^{\rm
cc}(d)=\left(\ket{00}\bra{00}+\ket{01}\bra{01}+\ket{10}\bra{10}\right)
\otimes\sigma_0+\ket{11}\bra{11}\otimes U(d), \eeq where $U(d)$ is a
diagonal or antidiagonal operation of one-qubit systems. Just as
well-known, it, together with the operations (\ref{co1}-\ref{co4}),
can be used to construct a universal gate.

The protocol proof of remote implementations of $N$-qubit operations
belonging to our restricted sets is given in Appendix B.

\section{Discussion and conclusion}\label{sec6}

In summary, we propose and prove the protocol of remote
implementations of partially unknown quantum operations of
multiqubits belonging to the restricted sets, and we obtain the
general and explicit forms of these restricted sets, that is, every
row and every column of an arbitrary member of operations belonging
to the restricted sets only has one nonzero element. Our protocol is
based on the simplified HPV scheme, but it can be thought of as a
development of HPV scheme to the cases of multiqubit systems since
our restricted sets of multiqubit operations are not simply
reducible to the direct products of HPV restricted sets of one-qubit
operations. Moreover, we have given evidence of the uniqueness and
optimization of our restricted sets based on the precondition that
our protocol only uses $N$ Bell's pairs. In order to show our
protocol in the above several aspects, we investigate in detail the
cases of two qubits. Note that those quantum operations with the
clearly physical significance and practical applications are
included in our restricted sets which can be remotely implemented.
It should be pointed out that the universal recovery operations
found by us are useful because they will be helpful for the design
of unified recovery quantum circuits in the near future. This
implies that the quantum operations that can be remotely implemented
are extended from only belonging to a given restricted set to
belonging to all of the restricted sets in our protocol. Its
advantages is obviously that the power of remote implementations of
quantum operations is enhanced. Of course, the unified recovery
operations need two mapping tables that are known, respectively, by
Alice and Bob before the beginning of the protocol.

In the area of resource consumption, the remote implementations of
quantum operations belonging to two restricted sets of one-qubit
operations need one $e$-bit which is shared by the sender and
receiver and three $c$-bits (or two when Bob fixes his preparing
way) from which one $c$-bit is transferred from the receiver to the
sender and two $c$-bits are transferred from the sender to the
receiver. In our protocol, we can see that the remote
implementations of quantum operations belonging to 24 restricted
sets of two-qubit operations need two $e$-bits and nine $c$-bits (or
seven $c$-bits when Bob fixes his preparing way), where two $e$-bits
are shared by the sender and the receiver, respectively, and two of
nine $c$-bits are transferred from the receiver to the sender while
the other seven $c$-bits are transferred from the sender to the
receiver. For the case of $N$ qubit operations, since the number of
restricted sets that can be remotely implemented is $2^N!$, their
remote implementations need $N$ $e$-bits and $2N+[\ln_2(2^N!)]+1$
$c$-bits (or $N+[\ln_2(2^N!)]+1$ $c$-bits when Bob fixes his
preparing way), where $N$ $e$-bits are shared by the sender and the
receiver, and $N$ $c$-bits are transferred from the receiver to the
sender while the other $N+[\ln_2(2^N!)]+1$ $c$-bits are transferred
from the sender to the receiver. Here, ``$[x]$" means taking the
integer part of $x$. In addition, the fixed local operations
$R_N(x)$ need to be used, and two mapping tables from $T^r_N(x,t)$
to a classical information $x$ and from a classical information $x$
to $R_N(x)$ need to be built before the beginning of the protocol.
Usually, the number of interesting restricted sets that can be
remotely implemented may be small, and the classical resource can be
correspondingly decreased. However, this will pay the price that the
power of protocol of remote implementations of quantum operations is
reduced. It should be pointed out that the implementations of
nonlocal quantum operations are different from the remote
implementations of the quantum operations. Therefore, the resource
used by them may be different in general.

Similar to the conclusion provided by Refs. \cite{Huelga1,Huelga2},
we have not found a faithful scheme without using the maximum
entanglement \cite{Our1}. Actually, this is partially because there
is no obvious physical significance when a unitary operation
belonging to the restricted sets acts on a density matrix of
diagonal state, and such an action is equivalent to the known one
that will be used in the recovery operation. For example, a phase
gate on one qubit acting on a density matrix of a diagonal state
gives nothing, an antidiagonal unitary transformation on one qubit
acting on a density matrix of a diagonal state is just a flip gate.
Of course, the study on the possible tradeoffs between the
entanglement and classical communication will still be important in
the near future.

Furthermore, we can investigate the controlled remote
implementations of partially unknown quantum operations belonging to
the restricted sets of one- and multiqubits. Similar to the
controlled teleportation of a quantum state via the GHZ states, the
controlled remote implementations of partially unknown quantum
operations can use the GHZ states which are a very important quantum
information resource \cite{GHZ}. In our view, the controlled remote
implementations of quantum operations should have some remarkable
applications in the remote quantum information processing and
communication including the future quantum internet. Here, a quantum
internet is a counterpart to the classical one, but it connects some
quantum computers that are located at different places together and
is used for the remote communication of quantum information and
remote implementations of quantum operations. The relevant
conclusions are studied in \cite{OurCCRIO}.

\section*{Acknowledgments}

We acknowledge all the collaborators of our quantum theory group at
the Institute for Theoretical Physics of our university. In
particular, we are grateful to Ning Bo Zhao, Xiao San Ma, Bo Sheng
Zhang, Cheng Zhao and Dong Zheng for their suggestive discussions.
This work was funded by the National Fundamental Research Program of
China under No. 2001CB309310, and partially supported by the
National Natural Science Foundation of China under Grant No.
60573008.

\begin{appendix}

\section{Swapping transformation}

Here, we study the general swapping transformations, which are
combinations of a series of usual swapping transformations. They are
used in our protocol in order to express our formula clearly and
compactly, and prove our protocol easily and strictly.

Note that a swapping transformation of two neighbor qubits ($2\times
2$ matrix) is defined by \beq S_W=\left(\begin{array}{cccc}
1 & 0 & 0 &0\\
0& 0 & 1 & 0\\
0& 1 & 0 & 0\\
0& 0 & 0 & 1
\end{array}\right).
\eeq Its action is \beq
S_W\ket{\alpha_X\beta_Y}=\ket{\beta_Y\alpha_X},\quad S_W(M^X\otimes
M^Y)S_W=M^Y\otimes M^X. \eeq This means that the swapping
transformation changes the space structure $H_X\otimes H_Y$ into
$H_Y\otimes H_X$.

For an $N$-qubit system, the swapping gate of the $i$th qubit and
the $i+1$th qubit reads \beq S_N(i,i+1)=
\sigma_0^{\otimes(i-1)}\otimes S_W\otimes\sigma_0^{\otimes(N-i-1)}.
\eeq Two rearranged transformations are defined by
\begin{equation}
F_N(i, j)=\prod_{\alpha=1\leftarrow}^{j-i}S_N(j-\alpha,j+1-\alpha)
\end{equation}
\begin{equation}
P_N(j, k)=\prod_{\beta=j\leftarrow}^{k-1}S_N(\beta,\beta+1)
\end{equation}
where $F_N(i,j)$ extracts out the spin-state of site $j$, and
 rearranges it forwards to the site $i$ ($i<j$) in the qubit-string,
where $P_N(j,k)$ extracts out the spin-state of site $j$, and
rearranges it backwards to the site $k$ ($k>j$) in the qubit-string.
Note that ``$\leftarrow$" means that the factors are arranged from
right to left corresponding to $\alpha, \beta$ from small to large.
Now, in terms of $P(j,k)$, we can introduce two general swapping
transformations with the forms
\begin{equation}
{\Lambda}(2,N)=\prod_{i=1\leftarrow}^{N-1}P_{2N}\left(2(N-i),2N-i\right),\quad
(N\geq 2)
\end{equation}
\begin{equation}
{\Omega}(2,N)=\prod_{i=1\leftarrow}^{N}P_{2N}\left(1,2N\right),\quad
(N\geq 2)
\end{equation}
Thus, \beq {\Lambda}(2,N)\left(\bigotimes_{i=1}^N\ket{a_i
b_i}\right)=\left(\bigotimes_{i=1}^N
\ket{a_i}\right)\otimes\left(\bigotimes_{j=1}^N
\ket{b_j}\right),\eeq \beq {\Lambda}(2,N)\left(\bigotimes_{k=1}^N
\left(M_{\alpha_i}^{A_i}\otimes
M_{\beta_i}^{B_i}\right)\right){\Lambda}^{-1}(2,N)=\left(\bigotimes_{i=1}^N
M_{\alpha_i}^{A_i}\right)\otimes\left(\bigotimes_{j=1}^N
M_{\beta_j}^{B_j}\right), \eeq \beq
{\Omega}(2,N)\left[\left(\bigotimes_{i=1}^N\ket{a_i
}\right)\otimes\left(\bigotimes_{j=1}^N\ket{b_j
}\right)\right]=\left(\bigotimes_{i=1}^N
\ket{b_i}\right)\otimes\left(\bigotimes_{j=1}^N
\ket{a_j}\right),\eeq \beq
{\Omega}(2,N)\left[\left(\bigotimes_{i=1}^N
M_{\alpha_i}^{A_i}\right)\left(\bigotimes_{i=1}^N
M_{\beta_i}^{B_i}\right)\right]{\Omega}^{-1}(2,N)=\left(\bigotimes_{i=1}^N
M_{\beta_i}^{B_i}\right)\otimes\left(\bigotimes_{j=1}^N
M_{\alpha_j}^{A_j}\right). \eeq Similarly, we can introduce
\begin{equation}
{\Upsilon}(3,N)=\prod_{i=1\leftarrow}^{N-1}F_{3N}\left(3i,2N+i\right),\quad
(N\geq 2).
\end{equation}
\beq \Gamma(3,N)=\left(I_{2^N}\otimes
\Omega(2,N)\right)\left(\Lambda(2,N)\otimes I_{2^N}\right).\eeq
Thus, \beq {\Upsilon}(3,N)\left(\bigotimes_{i=1}^N\ket{a_i
b_i}\right)\otimes\left(\bigotimes_{j=1}^N\ket{y_j}\right)=\bigotimes_{i=1}^N
\ket{a_i b_i y_i},\eeq \beq {\Upsilon}(3,N)\left[\bigotimes_{k=1}^N
\left(M_{\alpha_i}^{A_i}\otimes
M_{\beta_i}^{B_i}\right)\right]\left(\bigotimes_{j=1}^N
M_{\gamma_j}^{Y_j}\right){\Upsilon}^{-1}(3,N) =\bigotimes_{i=1}^N
M_{\alpha_i}^{A_i}\otimes M_{\beta_i}^{B_i}\otimes
M_{\gamma_i}^{Y_i},\eeq \beq
{\Gamma}(3,N)\left(\bigotimes_{i=1}^N\ket{a_i
b_i}\right)\otimes\left(\bigotimes_{j=1}^N\ket{y_j}\right)=\left(\bigotimes_{i=1}^N
\ket{a_i}\right)\otimes\left(\bigotimes_{j=1}^N\ket{
y_j}\right)\otimes\left(\bigotimes_{k=1}^N \ket{b_k}\right),\eeq
\beqa & & {\Gamma}(3,N)\left[\bigotimes_{k=1}^N
\left(M_{\alpha_i}^{A_i}\otimes
M_{\beta_i}^{B_i}\right)\right]\left(\bigotimes_{j=1}^N
M_{\gamma_j}^{Y_j}\right){\Gamma}^{-1}(3,N)\nonumber\\
& &\quad =\left(\bigotimes_{i=1}^N
M_{\alpha_i}^{A_i}\right)\otimes\left(\bigotimes_{j=1}^N
M_{\gamma_j}^{Y_j}\right)\otimes\left(\bigotimes_{k=1}^N
M_{\beta_k}^{B_k}\right). \eeqa

More generally, consider the set $\mathbb{Q}_N$ to be a whole
permutation of the bit-string ${a_1a_2\cdots a_N}$, and denote the
$z$th element with a bit-string form $Q(z)=q_1(z)q_2(z)\cdots
q_N(z)$, we can always obtain such a general swapping transformation
$W_N$ that a computational basis $\ket{a_1a_2\cdots a_N}$ of
$N$-qubit systems can be swapped as another basis
$\ket{q_1(z)q_2(z)\cdots q_N(z)}$ in which $q_1(z)q_2(z)\cdots
q_N(z)$ is an arbitrary element of $\mathbb{Q}_N$. That is, we can
write a given general swapping transformation $W_N\left[a_1a_2\cdots
a_N\rightarrow q_1(z)q_2(z)\cdots q_N(z)\right]$, \beq
W_N\left[{a_1a_2\cdots a_N}\rightarrow {q_1(z)q_2(z)\cdots
q_N(z)}\right]\ket{a_1a_2\cdots a_N}=\ket{q_1(z)q_2(z)\cdots
q_N(z)}.\eeq Furthermore, if we denote two dimensional space $A_i$
spanned by $\ket{a_i}$ ($a_i=0,1$ and $i=1,2,\cdots N$), while
$M^{A_i}$ is a matrix belonging to this space, we obviously have
\beqa & &W_N^{-1}\left[{a_1a_2\cdots a_N}\rightarrow
{q_1(z)q_2(z)\cdots
q_N(z)}\right]\left(\prod_{i=1}^NM^{A_i}\right)\nonumber\\
& & W_N\left[{a_1a_2\cdots a_N}\rightarrow {q_1(z)q_2(z)\cdots
q_N(z)}\right]=\left(\prod_{i=1}^NM^{A_{q_i(z)}}\right).\eeqa
Therefore, the general swapping transformation $W_N$ defined above
can be used to change the space structure of multiqubits systems.

\section{Proof of our protocol}

Here, we would like to prove our protocol of remote implementations
of quantum operations belonging to our restricted sets in the cases
with more than one qubit.

By using the swapping transformation $\Upsilon$, we can rewrite the
initial state \beq \ket{\Psi^{\rm
ini}_N}=\frac{1}{\sqrt{2^N}}\Upsilon^{-1}(3,N)\sum_{k_1,\cdots
k_N=0}^1 y_{k_1\cdots
k_N}\bigotimes_{m=1}^N\left(\ket{00k_m}+\ket{11k_m}\right). \eeq
From Bob's preparation, it follows that \beqa \ket{\Psi^P(b_1,\cdots
b_N)}&=&\mathcal{P}_B(b_1,b_2,\cdots,b_N)\ket{\Psi^{\rm
ini}_N}\nonumber\\
&=&\frac{1}{\sqrt{2^N}}\Upsilon^{-1}(3,N)\sum_{k_1,\cdots k_N=0}^1
y_{k_1\cdots k_N}\nonumber\\
& &\bigotimes_{m=1}^N \left\{\sigma_0\otimes
\left[\left(\ket{b_m}\bra{b_m}\right) C^{\rm
not}(0,1)\right]\right\}\left[
\left(\ket{00k_m}+\ket{11k_m}\right)\right]. \eeqa

Note that \beqa  & &\left\{\sigma_0\otimes
\left[\left(\ket{b}\bra{b}\right) C^{\rm
not}(0,1)\right]\right\}\left[
\left(\ket{00k}+\ket{11k}\right)\right]\nonumber\\
&=&\left[\sigma_0\otimes \left(\ket{b}\bra{b}\right)
\otimes\sigma_0\right]
\left\{\left(\ket{000}+\ket{110}\right)\delta_{k0}
+\left(\ket{011}+\ket{101}\right)\delta_{k1}\right\}\nonumber\\
&=&\left(\ket{0b0}\delta_{b0}+\ket{1b0}\delta_{b1}\right)\delta_{k0}
+\left(\ket{0b1}\delta_{b1}+\ket{1b1}\delta_{b0}\right)\delta_{k1}\nonumber\\
&=& \left[ \ket{bb0}\left(\delta_{b0}+\delta_{b1}\right)\delta_{k0}
+\ket{(1-b)b1}\left(\delta_{b1}+\delta_{b0}\right)\delta_{k1}\right]\nonumber\\
&=&\left(\sigma_{b}\otimes I_4\right)\left(\delta_{k0}\ket{0b0}
+\delta_{k1}\ket{1b1}\right)
\nonumber\\
&=&\left(\sigma_{b}\otimes
I_4\right)\left(\delta_{k0}+\delta_{k1}\right)\ket{kbk}\nonumber\\
&=& \left(\sigma_{b}\otimes I_4\right)\ket{kbk}, \eeqa where we have
used the facts that $\sigma_b\ket{b}=\ket{0}$ and
$\sigma_b\ket{1-b}=\ket{1}$ for $b=0,1$. This results in \beqa
\ket{\Psi^P(b_1,\cdots b_N)}&=&
\frac{1}{\sqrt{2^N}}\Upsilon^{-1}(3,N)\sum_{k_1,\cdots k_N=0}^1
y_{k_1\cdots k_N}\bigotimes_{m=1}^N
\left(\sigma_{b_m}\otimes\sigma_0\otimes\sigma_0\right)\ket{k_mb_mk_m}\nonumber\\
&=&\frac{1}{\sqrt{2^N}}\left[\bigotimes_{m=1}^N
\left(\sigma_{b_m}\otimes\sigma_0\right)\otimes
I_{2^N}\right]\sum_{k_1,\cdots k_N=0}^1 y_{k_1\cdots
k_N}\bigotimes_{m=1}^N\ket{k_mb_mk_m}\otimes\bigotimes_{m=1}^N\ket{k_m}
\nonumber\\
&=&\frac{1}{\sqrt{2^N}}\left[\bigotimes_{m=1}^N
\left(\sigma_{b_m}\otimes\sigma_0\right)\otimes
I_{2^N}\right]\Gamma_N^{-1}\sum_{k_1,\cdots k_N=0}^1 y_{k_1\cdots
k_N}\nonumber\\
& &\bigotimes_{m=1}^N\ket{k_m}\otimes
\bigotimes_{m=1}^N\ket{k_m}\otimes\bigotimes_{m=1}^N\ket{b_m}, \eeqa
where $\Gamma_N$ is defined by \beq \Gamma_N=\left(I_{2^N}\otimes
\Omega(2,N)\right)\left(\Lambda(2,N)\otimes I_{2^N}\right),\eeq
while $\Lambda(2,N)$ and $\Omega(2,N)$ are defined in Appendix A.

After Alice's sending and Bob's recovery operation, we have \beqa
\ket{\Psi^{\rm final}_N(x)}&=&
\frac{1}{\sqrt{2^N}}\Gamma_N^{-1}\sum_{k_1,\cdots k_N=0}^1
y_{k_1\cdots k_N}\left(\bigotimes_{m=1}^N
\ket{a_m}_{A_m}\right)\nonumber\\
& & \left[\left(\bigotimes_{m=1}^N \bra{a_m}\right)\cdot
\left(\bigotimes_{m=1}^N H^{A_m}\right)\cdot
T^r_N(x)\left(\bigotimes_{m=1}^N \ket{k_m}\right)\right]\nonumber\\
& &\otimes\left[\left(\bigotimes_{m=1}^N
\mathfrak{r}(a_m)\right)\cdot
{R}_N(x)\left(\bigotimes_{m=1}^N\ket{k_m}_{Y_m}\right)\right]
\otimes\left(\bigotimes_{m=1}^N\ket{b_m}_{B_m}\right). \eeqa

Thus, Alice's sending step and Bob's recovery operations yield the
final state in our interesting subsystem as \beqa \label
{nqfs1}\ket{\Psi^{\rm final}_N(x)}&=&
\frac{1}{\sqrt{2^N}}\Gamma_N^{-1}
\bigotimes_{m=1}^N\ket{a_m}_{A_m}\otimes\left\{\sum_{k_1,\cdots
k_N=0}^1 y_{k_1\cdots k_N}\right.\nonumber\\
& & \left[\left(\bigotimes_{m=1}^N
\bra{a_m}\right)\left(\bigotimes_{m=1}^N
H\right)T_N^r(x,t)\left(\bigotimes_{n=1}^N
\ket{k_n}\right)\right]\nonumber\\
& &\left.
\left(\bigotimes_{m=1}^N\mathfrak{r}^{Y_m}(a_m)\right)R_N(x)
\left(\bigotimes_{m=1}^N\ket{k_m}_{Y_m}\right)\right\}
\otimes\left(\bigotimes_{m=1}^N\ket{b_m}_{Y_m}\right).\eeqa

It is a key matter that we can prove the relation \beq \label{ppr1}
T_N^r(1,t)R_N(x)=\sum_{m=1}^{2^N}
t_m\ket{m,D}\bra{m,D}\sum_{n=1}^{2^N}\ket{n,D}\bra{p_n(x),D}
=\sum_{m=1}^{2^N}t_m\ket{m,D}\bra{p_m(x),D}=T_N^r(x,t).\eeq
According to the translation from the binary system to the decimal
system, we can rewrite $t_m$ as $t_{j_1\cdots j_N}$. So, the
diagonal $T^r_N(1,t)$ becomes \beq
T^r_N(1)=\sum_{j_1,\cdots,j_N=0}^{1}t_{j_1j_2\cdots
j_N}\ket{j_1j_2\cdots j_N}\bra{j_1j_2\cdots j_N}.\eeq In addition,
we know \beq \mathfrak{r}(a_m)=\sum_{l_m=0}^1(-1)^{a_m
l_m}\ket{l_m}\bra{l_m}.\eeq Substituting them into (\ref{nqfs1}), we
have \beqa \label{nqfs2}\ket{\Psi^{\rm final}_N(x)}&=&
\frac{1}{\sqrt{2^N}}\Gamma_N^{-1}
\bigotimes_{m=1}^N\ket{a_m}_{A_m}\otimes\left\{\sum_{j_1,\cdots,j_N=0}^1
\sum_{k_1,\cdots k_N=0}^1\sum_{l_1,\cdots,l_N} t_{j_1\cdots j_N}
y_{k_1\cdots
k_N}\right.\nonumber\\
& &\times\left(\prod_{m=1}^N\bra{a_m}H\ket{j_m}\right)
\left[\left(\bigotimes_{i=1}^N\bra{j_m}\right)R_N(x)
\left(\bigotimes_{n=1}^N\ket{k_n}\right)\right]\nonumber\\
& &\times\left[\left(\bigotimes_{m=1}^N\bra{l_m}\right)R_N(x)
\left(\bigotimes_{n=1}^N\ket{k_n}\right)\right]
\left(\prod_{m=1}^N(-1)^{a_m
l_m}\right)\nonumber\\
& &\left(\bigotimes_{m=1}^N\ket{l_m}_{Y_m}\right)
\otimes\left(\bigotimes_{m=1}^N\ket{b_m}_{B_m}\right).\eeqa Because
that $R_N(x)$ is such a matrix that its every row and every column
only has one nonzero element and its value is 1, we can obtain \beqa
& &\left[\left(\bigotimes_{m=1}^N\bra{j_m}\right)R_N(x)
\left(\bigotimes_{m=1}^N\ket{k_m}\right)\right]\left[
\left(\bigotimes_{m=1}^N\bra{l_m}\right)R_N(x)
\left(\bigotimes_{m=1}^N\ket{k_m}\right)\right]
\nonumber\\
& & =\left(\prod_{m=1}^N\delta_{j_m l_m}\right)
\left[\left(\bigotimes_{m=1}^N\bra{j_m}\right)R_N(x)
\left(\bigotimes_{m=1}^N\ket{k_m}\right)\right]. \eeqa Again from
\beq \bra{a_m}H\ket{j_m}(-1)^{a_m j_m}=\frac{1}{\sqrt{2}},\eeq we
can derive \beqa \label{nqfs3}\ket{\Psi^{\rm final}_N(x)}&=&
\frac{1}{\sqrt{2^N}}\Gamma_N^{-1}
\bigotimes_{m=1}^N\ket{a_m}_{A_m}\otimes\left\{\sum_{j_1,\cdots,j_N}\sum_{k_1,\cdots
k_N=0}^1\sum_{l_1,\cdots,l_N} t_{j_1\cdots j_N} y_{k_1\cdots
k_N}\right.\nonumber\\
& &\left[\left(\bigotimes_{m=1}^N\bra{j_m}\right)R_N(x)
\left(\bigotimes_{m=1}^N\ket{k_m}_{Y_m}\right)\right]
\left(\bigotimes_{m=1}^N\ket{j_m}_{Y_m}\right)\otimes
\left(\bigotimes_{m=1}^N\ket{b_m}_{B_m}\right).\eeqa

If we directly act $T_N^r(x,t)$ on the unknown state, we have \beqa
T^r_N(x,t)\ket{\xi}_{k_1\cdots k_N}&=&\sum_{k_1,\cdots,k_N=0}^1
y_{k_1\cdots k_N}T^r_N(1,t)R(x)\ket{k_1k_2\cdots k_n}\nonumber\\
&=&\sum_{j_1,\cdots,j_N=0}^1\sum_{k_1,\cdots,k_N=0}^1 t_{j_1\cdots
j_N}y_{k_1\cdots k_N}\bra{j_1j_2\cdots j_N}R(x)\ket{k_1k_2\cdots
k_n}\nonumber\\
& &\ket{j_1j_2\cdots j_N}.\eeqa This means that \beq
\label{nqfs4}\ket{\Psi^{\rm final}_N(x)}= \frac{1}{\sqrt{2^N}}
\bigotimes_{m=1}^N\ket{a_mb_m}_{A_mB_m}\otimes \left(
T^r_N(x,t)\ket{\xi}_{Y_1\cdots Y_N}\right).\eeq Here, we have
restored the structure of Hilbert's space by dropping the swapping
transformations. Therefore, we finish the proof of our protocol of
remote implementations of $N$-qubit operations belonging to our
restricted sets.

\end{appendix}

\end{document}